\documentclass[journal]{IEEEtran}

\usepackage{cite}
\usepackage{graphicx}

\usepackage[cmex10]{amsmath}

\usepackage{verbatim}
\usepackage{relsize}
\usepackage{amsfonts}
\usepackage{amsthm}
\usepackage{amssymb}
\usepackage{latexsym}
\usepackage{url}
\usepackage{epstopdf}
\usepackage{enumerate}
\usepackage{mathrsfs}
\usepackage{tabulary}
\usepackage{color}
\usepackage{balance}

\usepackage[justification=centering]{caption}

\newtheorem{theorem}{Theorem}
\newtheorem{lemma}{Lemma}
\newtheorem{definition}{Definition}

\newtheorem{proposition}{Proposition}

\newcommand{\floor}[1]{\left\lfloor #1\right\rfloor}
\newcommand{\ceil}[1]{\left\lceil #1\right\rceil}

\def\myfigwidth{0.5}

\begin{document}
\title{Are RLL Codes Suitable for Simultaneous Energy and Information Transfer?}
\author{Anshoo~Tandon,~\IEEEmembership{Member,~IEEE,} Mehul~Motani,~\IEEEmembership{Fellow,~IEEE},\\
	and Lav~R.~Varshney,~\IEEEmembership{Senior Member,~IEEE}
	\thanks{A. Tandon is with the Department of Computer Science, National University of Singapore, Singapore 117417 (email: anshoo.tandon@gmail.com)}%
	\thanks{M. Motani is with the Department of Electrical and Computer Engineering, National University of Singapore, Singapore 117583 (email: motani@nus.edu.sg).}%
	\thanks{L. R. Varshney is with the Coordinated Science Laboratory and the Department of Electrical and Computer Engineering, University of Illinois at Urbana-Champaign, Urbana, IL~61801 USA (email: varshney@illinois.edu)}} 

\maketitle

\begin{abstract}
Run-length limited (RLL) codes are a well-studied class of constrained codes having application in diverse areas such as optical and magnetic data recording systems, DNA-based storage, and visible light communication. RLL codes have also been proposed for the emerging area of simultaneous energy and information transfer, where the receiver uses the received signal  for decoding information as well as for harvesting energy to run its circuitry. In this paper, we show that RLL codes are \emph{not} the best codes for simultaneous energy and information transfer, in terms of the maximum number of codewords which avoid energy outage, i.e., outage-constrained capacity. Specifically, we show that sliding window constrained (SWC) codes and subblock energy constrained (SEC) codes have significantly higher outage-constrained capacities than RLL codes. 
\end{abstract}


\section{Introduction}
We consider simultaneous energy and information transfer from a powered transmitter to a receiver which relies completely on the received information-bearing signal for its real-time power requirements. In this scenario, the problem at the transmitter is to design codes which maximize information transfer rate while constraining codewords to carry sufficient energy content which powers the receiver circuitry. When on-off keying is employed, where ``1" (resp. ``0") is represented by the presence (resp. absence) of a carrier, a majority transmission of ``1" indicates a greater opportunity for the receiver to fulfill its power requirements. In order to meet the real-time energy requirement at the receiver, the use of \emph{run-length limited} (RLL) codes with sufficiently high runs of ``1" has been proposed~\cite{Barbero11,Rosnes12,Barbero14,Fouladgar14,Tandon14_ITA}. In this paper, we show that although RLL codes are an important class of constrained codes with diverse applications~\cite{ImminkBook91,Tang69,Franaszek70,Franaszek72,Siegel85,Siegel87,Zehavi88,Shamai90,ImminkBook99,Ghaffar91,Wijngaarden01,Yogesh06,Wijngaarden10,Cai17,Immink18,Uday18}, they are \emph{not} the most suitable class of constrained codes for simultaneous energy and information transfer in terms of outage-constrained capacity.  

RLL codes have a long and rich history and have been used in numerous applications~\cite{ImminkBook91}. An early work discussed the application of RLL coding techniques to magnetic recording~\cite{Tang69}. The use of finite state machine methods for analyzing RLL sequence constraint was first discussed by Franaszek~\cite{Franaszek70}. Efficient implementation of RLL codes, with error propagation limitation mechanism, was proposed in~\cite{Franaszek72}. An early tutorial introduction to RLL codes, their application to magnetic storage, and techniques for RLL code design and implementation was presented in~\cite{Siegel85}. The asymptotic capacity of RLL codes, under different channel conditions, have also been well studied~\cite{Siegel87,Zehavi88,Shamai90}. RLL codes have been applied in practice to various data storage devices, including virtually all magnetic and optical disc recording systems~\cite{ImminkBook99}. Over the years, different construction schemes for RLL codes, with varied enhancements, have been proposed and analyzed~\cite{Ghaffar91,Wijngaarden01,Yogesh06,Wijngaarden10}. Study of RLL codes has continued to be an important research topic, and recent work includes its application to high density data storage \cite{Cai17}, DNA-based storage \cite{Immink18}, and visible light communication~\cite{Uday18}.

The use of RLL codes for simultaneous energy and information transfer over an inductive coupled RFID channel was proposed in~\cite{Barbero11,Rosnes12,Barbero14}, where the tradeoff between capacity and energy content for certain specific RLL codes was discussed. 
A detailed analysis of the interplay between information rate and energy transfer using RLL codes for simultaneous energy and information transfer was investigated in~\cite{Fouladgar14}. It was shown that the parameters of RLL codes could be appropriately chosen based on the desired probabilities of battery underflow and overflow at the receiver. In~\cite{Tandon14_ITA}, the use of RLL codes with at least $d$ ones between successive zeros was proposed for simultaneous energy and information transfer, and capacity bounds were presented for different noisy channels.
	
In this paper, we show that, although RLL codes are an important class of constrained codes, they are \emph{not} the most suitable class of constrained codes for simultaneous energy and information transfer.  In particular, we show that other classes of constrained codes, such as the \emph{sliding window constrained} (SWC) codes~\cite{Rosnes12,Tandon14_ITA} and the \emph{subblock energy constrained} (SEC) codes~\cite{Tandon16_ISIT,Tandon17_SECC_ISIT}, provide significantly higher outage-constrained capacity compared to RLL codes.

SWC codes require that the number of ones within a sliding time window of fixed length exceed a certain threshold. The use of SWC codes for simultaneous energy and information transfer was discussed in~\cite{Rosnes12,Tandon14_ITA}. SEC codes, on the other hand, require that the number of ones within non-overlapping subblocks exceed a threshold. The use of SEC codes for real-time simultaneous energy and information transfer was proposed in \cite{Tandon16_CSCC_TIT}, and their properties have been analyzed in~\cite{Tandon16_CSCC_TIT,Tandon16_ISIT,Tandon17_SECC_ISIT,TandonKM17_Arxiv}. In this paper, we compare the noiseless capacity of these different classes of constrained codes under a common requirement of powering a receiver to ensure that it never suffers an energy outage, where an outage is an event triggered due to underflow of the energy buffer (battery). 

Practical applications of energy and information transfer include implantable biomedical devices~\cite{Yakovlev12,Yilmaz13}, and low-power Internet of Things (IoT) devices~\cite{Gollakota14}, which receive energy and control signals wirelessly from an external unit. Other examples include powerline communication where information is sent over the same lines which carry electric power~\cite{Pavlidou03}. Early work analyzing the fundamental tradeoff between energy and information transfer, in an information theoretic setting, was conducted in~\cite{Lav08} where codewords were constrained to have average receive energy exceed a threshold. Since then, there have been numerous extensions trading capacity and power under various settings~\cite{Grover10,Lav12,Popovski13,Ding15}.

The main contributions of this paper are as follows. 
\begin{itemize}
\item Firstly, we derive several bounds, relations, and properties of the noiseless capacity of RLL, SWC, and SEC codes (Sec.~\ref{Sec:Properties}). 
\item Secondly, we provide necessary and sufficient conditions on code parameters for avoiding energy outage at the receiver (Sec.~\ref{Sec:Avoid_Outage}). 
\item Thirdly, we use the results in Sec.~\ref{Sec:Properties} and Sec.~\ref{Sec:Avoid_Outage} to derive bounds and characterize the outage-constrained capacity of RLL, SWC, and SEC codes (Sec.~\ref{Sec:Capacity_under_no_outage}).
The main result of the paper is presented in Sec.~\ref{subsec:CompareOCC} where we prove that the outage-constrained capacity of RLL codes is just a lower bound to the outage-constrained capacity of SWC and SEC codes.
\item Finally, we provide numerical examples (Sec.~\ref{Sec:Numerical}) highlighting the impact of the receiver energy requirement and the energy buffer size on the outage-constrained capacity. The numerical results demonstrate that SWC and SEC codes provide significantly higher outage-constrained capacity than RLL codes.
\end{itemize}
  
We begin with a formal definition of different classes of constrained codes and their properties.

\section{Constrained sequences} \label{Sec:Notation}
For a real number $x$, let $\floor{x}$ denote the largest integer less than or equal to $x$, and $\ceil{x}$ denote the smallest integer greater than or equal to $x$. The set of positive integers is denoted $\mathbb{Z}^+$.

We now provide formal definitions of noiseless code capacity for RLL, SWC, and SEC codes.

\subsection{Run-length limited (RLL) sequences}
\begin{definition}
	Let $k \in \mathbb{Z}^+$ and $d \in \mathbb{Z}^+$ be such that $d < k$. A binary sequence is said to be a \emph{type-1 $(d,k)$-RLL sequence} if the number of ones between successive zeros are at least $d$ and at most $k$. Further, we let $k=\infty$ when there is no upper limit on the number of ones between successive zeros.
\end{definition}
Let $M_{RLL}^{(d,k)}(n)$ denote the number of distinct type-1 $(d,k)$-RLL sequences of length $n$. Then the noiseless code capacity using $(d,k)$-RLL sequences is defined as
\begin{equation}
C_{RLL}^{(d,k)} \triangleq \lim_{n \to \infty} \frac{\log M_{RLL}^{(d,k)}(n)}{n} . \nonumber
\end{equation}

Note that a type-1 $(d,k)$-RLL sequence is also a type-1 $(d,k+1)$-RLL sequence, as the $(d,k)$ constraint is stricter than the $(d,k+1)$ constraint. When RLL sequences are used for simultaneous energy and information transfer for avoiding energy outage at the receiver, potentially higher rates can be achieved by letting $k=\infty$. In the following, we will restrict attention to \emph{type-1} RLL sequences, and study $C_{RLL}^{(d,\infty)}$, the noiseless capacity using $(d,\infty)$-RLL sequences.

\subsection{Sliding window constrained (SWC) sequences}
\begin{definition}
	Let $T \in \mathbb{Z}^+$ and $w \in \mathbb{Z}^+$ be such that $w \le T$. A binary sequence $(b_1 b_2 \ldots b_n)$ is said to be a \emph{$(T,w)$-SWC sequence} if it satisfies the following constraint
	\begin{equation}
	\sum_{i = j+1}^{j + T} b_i \ge w , ~~j = 0, 1, \ldots, n-T. \nonumber
	\end{equation}
\end{definition}
The above constraint implies that the number of ones in a sliding window of length $T$ should be \emph{at least} $w$.

Let $M_{SWC}^{(T,w)}(n)$ denote the number of distinct $(T,w)$-SWC sequences of length $n$. Then the noiseless code capacity using $(T,w)$-SWC sequences is defined as
\begin{equation}
C_{SWC}^{(T,w)} \triangleq \lim_{n \to \infty} \frac{\log M_{SWC}^{(T,w)}(n)}{n} . \nonumber
\end{equation}

\subsection{Subblock energy constrained (SEC) sequences}
\begin{definition}
Let $L \in \mathbb{Z}^+$ and $w \in \mathbb{Z}^+$ be such that $w \le L$. A binary sequence $(b_1 b_2 \ldots b_n)$, where length $n$ is a multiple of $L$, is said to be an \emph{$(L,w)$-SEC sequence} if it satisfies the following constraint
\begin{equation}
\sum_{i = jL + 1}^{jL + L} b_i \ge w , ~~j = 0, 1, \ldots, (n/L)-1. \nonumber
\end{equation}
\end{definition}
The above constraint implies that if we partition the sequence into equal-sized subblocks of length $L$, then each subblock must have \emph{at least} $w$ ones.  Note that any $(T,w)$-SWC sequence is also a $(T,w)$-SEC sequence, as the sliding window constraint is stricter than the subblock constraint.

Let $M_{SEC}^{(L,w)}(n)$ denote the number of distinct $(L,w)$-SEC sequences of length $n$. Then the noiseless code capacity using $(L,w)$-SEC sequences is defined as
\begin{equation}
C_{SEC}^{(L,w)} \triangleq \lim_{n \to \infty} \frac{\log M_{SEC}^{(L,w)}(n)}{n} . \nonumber
\end{equation}
Because each subblock of length $L$ has at least $w$ ones, it follows that the capacity of $(L,w)$-SEC sequences is
\begin{equation}
C_{SEC}^{(L,w)} = \frac{1}{L} \log_2 \left[\sum_{i=w}^L \binom{L}{i} \right]. \label{eq:SEC_Lw_capacity}
\end{equation}

\subsection{Properties of constrained sequences} \label{Sec:Properties}
In this section, we present inequalities relating the capacities of constrained codes for different parameters.  The proofs for all results in this section are in the appendix.

The following proposition shows that the code capacity of $(d,\infty)$-RLL sequences is a strictly decreasing function of $d$.
\begin{proposition}\label{prop:rll}
	\label{prop:RLL_cap_dec_with_d}
	$C_{RLL}^{(d,\infty)}$ is strictly decreasing in $d$.
\end{proposition}

The next proposition bounds the capacity of SWC codes.
\begin{proposition} \label{prop1}
	The code capacity of $(T,w)$-SWC codes, $C_{SWC}^{(T,w)}$, satisfies the following inequalities
	\begin{equation}
	C_{SWC}^{(T+m,w+m)} \le C_{SWC}^{(T,w)} \le C_{SWC}^{(Tm,wm)} , \label{eq:SWC_Ineq}
	\end{equation}
	where $m \in \mathbb{Z}^+$.
\end{proposition}

Note that the lower bound in \eqref{eq:SWC_Ineq} implies that for any $m \in \mathbb{Z}^+$, the capacity $C_{SWC}^{(w+m,w)}$ is a non-increasing function of $w$. Similarly, for $m \in \mathbb{Z}^+$, it can also be shown that
\begin{equation}
C_{SWC}^{(T,w+m)} \le C_{SWC}^{(T,w)} \le C_{SWC}^{(T+m,w)}. \label{eq:SWC_Ineq2}
\end{equation}

The relation between noiseless capacities of $(T,w)$-SWC codes and $(T,w)$-SEC codes is given as follows.
\begin{proposition} \label{prop:SWC_capacity_LB_using_SEC_capacity}
	We have the inequality
	\begin{equation}
	\frac{T}{T+w} C_{SEC}^{(T,w)} \le C_{SWC}^{(T,w)} \le C_{SEC}^{(T,w)} \label{eq:SWC_SEC_ineq}.
	\end{equation}
\end{proposition}

Next, we provide an alternate lower bound on $C_{SWC}^{(T,w)}$.
\begin{proposition} \label{prop:SWC_capacity_LB_using_SEC_capacity2}
	For $m \in \mathbb{Z}^+$, we have
			\begin{equation}
			C_{SWC}^{(T,w)} \ge \max\left\{C_{SEC}^{\left(T-1,\ceil{(T+w-2)/2}\right)},  C_{SEC}^{\left(\floor{T/(m+1)},\ceil{w/m}\right)} \right\} . \nonumber
			\end{equation}
\end{proposition}

The next theorem shows that the $(d,\infty)$-RLL constraint is equivalent to the $(d+1,d)$-SWC constraint; two constraints are \emph{equivalent} if they induce the same set of codes.
\begin{theorem}\label{thm:swc}
	The $(d,\infty)$-RLL constraint is equivalent to the $(d+1,d)$-SWC constraint.
\end{theorem}

The following proposition is immediate.
\begin{proposition} \label{prop:RLL_SWC_eq}
	We have $C_{SWC}^{(d+1,d)} = C_{RLL}^{(d,\infty)}$.
\end{proposition}

It can also be shown that for fixed $w \in \mathbb{Z}^+$ and $m > 1$, the $(w+m,w)$-SWC constraint is \emph{not equivalent} to a $(d,k)$-RLL constraint for any $(d,k)$ pair, because a sequence starting with $m$ consecutive zeros can be constructed to satisfy the $(w+m,w)$-SWC constraint.

The following lemma uses Prop.~\ref{prop:RLL_SWC_eq} to show that the upper bound on $C_{SWC}^{(T,w)}$ in \eqref{eq:SWC_SEC_ineq} is strict when $w=T-1$. 
\begin{lemma} \label{lemma:SWC_SEC_capacity}
	For $T>1$, we have the strict inequality 
	\begin{equation*}
	C_{SWC}^{(T,T-1)} < C_{SEC}^{(T,T-1)}.
	\end{equation*}
\end{lemma}

\begin{lemma} \label{lemma:SEC_capacity_one_zero}
$C_{SEC}^{(T,T-1)}$ is a strictly decreasing function of $T$.
\end{lemma}
Lemmas \ref{lemma:SWC_SEC_capacity} and \ref{lemma:SEC_capacity_one_zero} will be used in Sec.~\ref{subsec:CompareOCC} to show that the outage-constrained capacity of RLL codes is just a lower bound to the outage-constrained capacity of SWC and SEC codes.

The next section presents necessary and sufficient conditions on the constrained codes to avoid energy outage at the receiver. 

\section{Avoiding Energy Outage} \label{Sec:Avoid_Outage}
In this section, we present necessary and sufficient conditions on the parameters of RLL, SWC, and SEC codes to avoid energy outage at the receiver. 

The energy recharge and usage model is as follows. Each arrival of bit-1 (resp. bit-0) brings in one (resp. zero) units of energy at the receiver. It is assumed that the receiver needs $B$ units of energy per bit arrival for its operation, where $0 < B~<~1$. Let $E_{max}$ denote the receiver energy buffer size, and let $E(i)$ denote its energy level at the start of the $i$th channel use. If the transmitted bits sequence is denoted $b_1 b_2 \cdots$, then the energy update equation is
\begin{equation}
	E(i+1) = \min\{|E(i) + b_i - B|^+ , E_{max} \} , \nonumber
\end{equation}
where the notation $|z|^+$ denotes $\max\{0,z\}$. An energy \emph{outage} is said to occur if $E(i) + b_i < B$, which captures the event where receiver energy buffer underflows. An \emph{overflow} is said to occur if $E(i) + b_i - B > E_{max}$.

\begin{theorem} \label{thm:RLL_conditions}
	Energy outage is avoided, over the set of all $(d,\infty)$-RLL codes, if and only if:
	\begin{align}
	d &\ge \ceil{\frac{B}{1-B}}, \label{eq:RLL_OA_cond1}\\
	E_{max} &\ge E(1) \ge B, \label{eq:RLL_OA_cond2}
	\end{align}
	where $E(1)$ denotes the initial energy level.
\end{theorem}
\begin{IEEEproof}
	We first prove necessity. Let the codeword length be $n = m(d+1)$ with $m \in \mathbb{Z}^+$, and $(b_1 b_2 \ldots b_n)$ be a $(d,\infty)$-RLL sequence, where every zero is followed by \emph{exactly} $d$ ones. The total energy content in this sequence is $m d$ units while the total energy required at the receiver is $m(d+1)B$. Thus, an outage will happen if $d < (d+1)B$, irrespective of the initial energy level $E(1)$, with sufficiently large $m$. Thus a necessary condition for outage avoidance is $d(1-B) \ge B$, which is equivalently expressed by \eqref{eq:RLL_OA_cond1}. Condition~\eqref{eq:RLL_OA_cond2} is required to avoid outage for a $(d,\infty)$-RLL sequence which begins with a zero.
	
	We now prove sufficiency. We know that every bit-0 decreases the energy buffer level by $B$ units, while bit-1 increases the energy level by $1-B$ units. For a $(d,\infty)$-RLL sequence which satisfies conditions \eqref{eq:RLL_OA_cond1} and \eqref{eq:RLL_OA_cond2}, a decrease in energy by $B$ units is immediately followed by an increase of at least $d (1-B) \ge \ceil{B/(1-B)} (1-B) \ge B$ units.
\end{IEEEproof}

\begin{theorem} \label{thm:SWC_conditions}
	Energy outage is avoided, over the set of all $(T,w)$-SWC codes, if and only if:
	\begin{align}
		w &\ge \ceil{TB}, \label{eq:SWC_OA_cond1}\\
		E_{max} &\ge E(1) \ge (T-w)B, \label{eq:SWC_OA_cond2}
	\end{align}
	where $E(1)$ denotes the initial energy level.
\end{theorem}
\begin{IEEEproof}
	We first prove necessity. Consider a $(T,w)$-SWC sequence of length $n = m T$, with $m \in \mathbb{Z}^+$, where a sliding window of length $T$ has \emph{exactly} $w$ ones. The total energy content in this sequence is $m w$ units while the total energy required at the receiver is $m T B$. Thus, an outage will result if $w < TB$, irrespective of the initial energy level $E(1)$, when $m$ is sufficiently large. Hence it is necessary that $w \ge \ceil{TB}$. The condition $E(1) \ge (T-w)B$ is needed to avoid outage for a sequence which begins with $T-w$ zeros.
	
	We now prove sufficiency. If there is no energy overflow during the first $T$ bit arrivals, conditions \eqref{eq:SWC_OA_cond1} and \eqref{eq:SWC_OA_cond2} imply that there is no outage in this interval with a $(T,w)$-SWC sequence, and we have $E(T+1) \ge E(1) + w - TB \ge E(1)$. Then we can recursively apply this argument to show that there will be no outage over the entire $(T,w)$-SWC sequence provided there is no energy overflow. We next show, by contradiction, that there is no outage even if some energy is lost due to overflow. Let $i$ denote the channel-use index where outage occurs, and let $j$ denote the last index where overflow occurs, $j < i$. Then $E(j+1) = E_{max} \ge E(1)$, and there is no overflow between indices $j+1$ and $i$. Thus, if we shift the origin to index $j+1$ (from index $1$), we get a contradiction to the claim that an outage cannot occur with a $(T,w)$-SWC sequence provided no energy is lost in overflow.
\end{IEEEproof}

\begin{theorem} \label{thm:SEC_conditions}
	Energy outage is avoided, over the set of all $(L,w)$-SEC codes, if and only if:
	\begin{align}
	w &\ge \ceil{LB}, \label{eq:SEC_OA_cond1}\\
	E(1) &\ge (L-w)B, \label{eq:SEC_OA_cond2}\\
	E_{max} &\ge 2(L-w)B, \label{eq:SEC_OA_cond3}	
	\end{align}
	where $E(1)$ denotes the initial energy level.
\end{theorem}
\begin{IEEEproof}
	We first prove necessity. Consider an $(L,w)$-SEC sequence of length $n = m L$, with $m \in \mathbb{Z}^+$, where each subblock of length $L$ has \emph{exactly} $w$ ones. The total energy content in this sequence is $m w$ units while the total energy required at the receiver is $m T B$. Thus, an outage will result if $w < TB$, irrespective of the initial energy level $E(1)$, when $m$ is sufficiently large. Hence it is necessary that $w \ge \ceil{TB}$. The condition $E(1) \ge (T-w)B$ is needed to avoid outage for a sequence which begins with $T-w$ zeros. The condition $E_{max} \ge 2(L-w)B$ is required to avoid outage for an $(L,w)$-SEC sequence, $(b_1 b_2 \ldots b_{mL})$, where the first subblock has $b_1 = b_2 = \cdots = b_w = 1$ and $b_{w+1} = \cdots = b_L = 0$, while the second subblock has $b_{L+1} = b_{L+2} = \cdots = b_{L+L-w} = 0$ and $b_{2L-(w-1)} = \cdots = b_{2L} = 1$. In this case, an outage occurs during channel use index $2L-w$ if condition \eqref{eq:SEC_OA_cond3} is violated.
	
	We now show that conditions \eqref{eq:SEC_OA_cond1}, \eqref{eq:SEC_OA_cond2}, and \eqref{eq:SEC_OA_cond3} are sufficient for avoiding outage in the \emph{first subblock}, and that $E(L+1) \ge (L-w)B$. This in turn will prove the sufficiency of these conditions for avoiding outage over the entire sequence by recursively applying the same argument over all subblocks. If there is no energy overflow in the first subblock, then \eqref{eq:SEC_OA_cond2} ensures no outage, and we have
	\begin{align}
	E(L+1) &\ge E(1) + w - LB \nonumber\\
	       &\ge E(1) \ge (L-w)B.  \nonumber
	\end{align}
	On the other hand, if there is energy overflow at $i$th index in the first subblock, then $E(i+1) = E_{max}$, and we have
	\begin{align}
	E(L+1) &\ge E(i+1) - (L-w)B  \nonumber\\
	       &= E_{max} - (L-w)B  \nonumber\\
	       &\ge (L-w)B .  \nonumber 
	\end{align}
\end{IEEEproof}

\section{Outage-Constrained Capacity} \label{Sec:Capacity_under_no_outage}
In this section, we formulate and analyze the outage-constrained capacity for given values of $B$ (required energy per bit) and $E_{max}$ (receiver energy buffer size). We assume the following initial condition for the energy level at the receiver
\begin{equation}
E(1) = E_{max} , \nonumber
\end{equation}
implying a saturated energy buffer level (full battery) at the start of transmission. Note that the energy buffer can be filled up in at most $\ceil{E_{max}/(1-B)}$ channel uses via a preamble consisting of all ones. This use of a fixed-length preamble does not impact the noiseless code capacity which is computed under the scenario where the blocklength tends to infinity.

\subsection{Run-length Limited Codes}
Let $\mathcal{A}_{RLL}^{E_{max}}(B)$ denote the set of all $d \in \mathbb{Z}^+$ which satisfy \eqref{eq:RLL_OA_cond1}. Then, the outage-constrained capacity for RLL codes, for energy buffer size $E_{max}$ and required energy per bit $B$, denoted $\mathcal{O}_{RLL}^{E_{max}}(B)$, is defined as
\begin{equation}
\mathcal{O}_{RLL}^{E_{max}}(B) \triangleq \max_{d \in \mathcal{A}_{RLL}^{E_{max}}(B)} C_{RLL}^{(d,\infty)} . \label{eq:RLL_cap_def}
\end{equation}

Note that among the set of all feasible values of $d$ which avoid outage at the receiver, the outage-constrained capacity is achieved by that value of $d$ which yields maximum capacity.

\begin{proposition} \label{prop:RLL_capacity_compute}
	We have
	\begin{align}
	\mathcal{O}_{RLL}^{E_{max}}(B) &= 0, \mathrm{~~~~~~~~~~~~~~~~~~~when~~} E_{max} < B \nonumber \\
	\mathcal{O}_{RLL}^{E_{max}}(B) &= C_{RLL}^{(\ceil{B/(1-B)},\infty)}, \mathrm{~~when~~} E_{max} \ge B. \nonumber
	\end{align} 
\end{proposition}
\begin{IEEEproof}
	If $E_{max} < B$ then \eqref{eq:RLL_OA_cond2} is violated irrespective of the choice of $d$, and hence $\mathcal{O}_{RLL}^{E_{max}}(B) = 0$. For $E_{max} \ge B$, combining Prop.~\ref{prop:RLL_cap_dec_with_d} and~\eqref{eq:RLL_OA_cond1}, we see that the maximum in \eqref{eq:RLL_cap_def} is achieved when $d = \ceil{B/(1-B)}$.
\end{IEEEproof}
Hence, the exact value of $\mathcal{O}_{RLL}^{E_{max}}(B)$ can easily be computed as the logarithm of the largest real root of $X^{d+1} - X^d -1$ where $d = \ceil{B/(1-B)}$.

\subsection{Sliding Window Constrained Codes}
Let $\mathcal{A}_{SWC}^{E_{max}}(B)$ denote the set of all feasible pairs $(T,w)$ which satisfy conditions \eqref{eq:SWC_OA_cond1} and \eqref{eq:SWC_OA_cond2}, such that corresponding $(T,w)$-SWC codes avoid outage. Then, the outage-constrained capacity for SWC codes, for energy buffer size $E_{max}$ and required energy per bit $B$, denoted $\mathcal{O}_{SWC}^{E_{max}}(B)$, is defined as
\begin{equation}
\mathcal{O}_{SWC}^{E_{max}}(B) \triangleq \max_{(T,w) \in \mathcal{A}_{SWC}^{E_{max}}(B)} C_{SWC}^{(T,w)} . \label{eq:SWC_Emax_B_capacity}
\end{equation}
If $E_{max} < B$, then condition \eqref{eq:SWC_OA_cond2} is satisfied only when $w=T$, and hence $\mathcal{O}_{SWC}^{E_{max}}(B) = 0$ in this case. 

The following proposition gives a lower bound on the outage-constrained capacity of SWC codes.
\begin{proposition} \label{prop:SWC_capacity_LB}
Let $z = \floor{E_{max}/B}$. Then for $z>0$, the capacity $\mathcal{O}_{SWC}^{E_{max}}(B)$ is lower bounded by $C_{SWC}^{(T,w)}$ where $T = \ceil{z/(1-B)}$ and $w = \ceil{TB} = T-z$. Further, there exists a capacity achieving pair $(T_0,w_0) \in \mathcal{A}_{SWC}^{E_{max}}(B)$ which satisfies $T_0 \le \ceil{z/(1-B)}$ and $w_0 = \ceil{T_0 B}$.
\end{proposition}
\begin{IEEEproof}
Observe that conditions \eqref{eq:SWC_OA_cond1} and \eqref{eq:SWC_OA_cond2} are equivalent to the following properties:
\begin{enumerate}[(i)]
	\item The fraction of ones in a sliding window of length $T$ is at least $B$.
	\item The number of zeros in a sliding window of length $T$ is at most $\floor{E_{max}/B} = z$.
\end{enumerate}
If $T = \ceil{z/(1-B)}$ and the number of ones in a sliding window of size $T$ are exactly $\ceil{TB}$, then it can be verified that the number of zeros in this window are equal to $T - \ceil{TB} = z$ thereby satisfying the above properties, and thus this pair $(T,w)$ with $w=\ceil{TB}$ belongs to the feasible set $\mathcal{A}_{SWC}^{E_{max}}(B)$.

From \eqref{eq:SWC_Ineq2} we observe that for a fixed $T$, the capacity can only increase upon reducing $w$. Thus, it follows from property $\mathrm{(i)}$ above that a capacity achieving pair $(T_0,w_0)$ satisfies $w_0 = \ceil{T_0 B}$. Now, we know from \eqref{eq:SWC_Ineq} that capacity will only decrease if both $T$ and $w$ increase by the same amount. On the other hand, if an increase in $T$ beyond $\ceil{z/(1-B)}$ does not increase $w$ by an equal amount, then the number of zeros in a window exceed $z$, thereby violating the second property. Thus, $T_0$ can be upper bounded by $\ceil{z/(1-B)}$. 
\end{IEEEproof}
Prop.~\ref{prop:SWC_capacity_LB} shows that
\begin{equation}
\mathcal{O}_{SWC}^{E_{max}}(B) \ge C_{SWC}^{\left(\ceil{\frac{1}{1-B} \floor{\frac{E_{max}}{B}}}, \ceil{B \ceil{\frac{1}{1-B} \floor{\frac{E_{max}}{B}}}}\right)}. \label{eq:SWC_capacity_LB_explicit}
\end{equation}

Thus, a computationally efficient lower bound on $\mathcal{O}_{SWC}^{E_{max}}(B)$ can be obtained by combining \eqref{eq:SWC_capacity_LB_explicit} with lower bounds on $C_{SWC}^{(T,w)}$ in Prop.~\ref{prop:SWC_capacity_LB_using_SEC_capacity} and Prop.~\ref{prop:SWC_capacity_LB_using_SEC_capacity2}.

The next proposition gives an upper bound on the outage-constrained capacity of SWC codes.
\begin{proposition} \label{prop:SWC_Capacity_UpperBound}
	We have 
	\begin{equation}
	\mathcal{O}_{SWC}^{E_{max}}(B) \le h\left(\max\{B,0.5\}\right), \label{eq:SWC_Capacity_UpperBound}	
	\end{equation}
	where $h(\cdot)$ denotes the binary entropy function, $h(x) \triangleq -x\log_2(x) - (1-x)\log_2(1-x)$.
\end{proposition}
\begin{IEEEproof}
For avoiding outage, the fraction of ones in an SWC sequence should be at least $B$. Thus, outage avoiding SWC sequences are a subset of the set of sequences where the fraction of ones is at least $B$. The proposition follows because the noiseless capacity of sequences with at least $B$ ones is equal to $h\left(\max\{B,0.5\}\right)$.
\end{IEEEproof}
We remark that $h\left(\max\{B,0.5\}\right)$ is an upper bound to the outage-constrained capacity for \emph{any} class of constrained code because the fraction of ones in any outage avoiding sequence are at least $B$.

The following proposition shows that $\mathcal{O}_{SWC}^{E_{max}}(B)$ achieves the upper bound in \eqref{eq:SWC_Capacity_UpperBound} when the receiver energy buffer size is unbounded.
\begin{proposition} \label{prop:SEC_capacity_large_Emax}
We have
\begin{equation}
\lim_{E_{max} \to \infty} \mathcal{O}_{SWC}^{E_{max}}(B) = h\left(\max\{B,0.5\}\right) . \nonumber 
\end{equation}
\end{proposition}
\begin{IEEEproof}
Let $T = \ceil{\frac{1}{1-B} \floor{\frac{E_{max}}{B} } }$ and use \eqref{eq:SWC_capacity_LB_explicit} to obtain
\begin{align}
\lim_{E_{max} \to \infty} \mathcal{O}_{SWC}^{E_{max}}(B) &\ge \lim_{T \to \infty} C_{SWC}^{(T, \ceil{TB})} \nonumber \\
&\ge \lim_{T \to \infty} C_{SEC}^{\left(\floor{\frac{T}{m+1}}, \ceil{\frac{1}{m}\ceil{TB}}\right)} \label{eq:SWC_capacity_LB_temp1} ,
\end{align}
where \eqref{eq:SWC_capacity_LB_temp1} follows from Prop.~\ref{prop:SWC_capacity_LB_using_SEC_capacity2}. Letting $T$ grow as $k^2$ and setting $m = k-1$, we get
\begin{equation}
\lim_{E_{max} \to \infty} \mathcal{O}_{SWC}^{E_{max}}(B) \ge \lim_{k \to \infty} C_{SEC}^{\left(k, \ceil{\ceil{k^2 B}/(k-1)}\right)} . \nonumber 
\end{equation}
For $k>3$, we have $\ceil{\ceil{k^2 B}/(k-1)} < \ceil{k B} + 2$. Hence
\begin{align}
\lim_{E_{max} \to \infty} \mathcal{O}_{SWC}^{E_{max}}(B) &\ge \lim_{k \to \infty} C_{SEC}^{\left(k, \ceil{kB}+2\right)} \nonumber \\
&= \lim_{k \to \infty} \frac{1}{k} \log_2 \left[\sum_{i=\ceil{k B}+2}^k \binom{k}{i} \right] \nonumber \\
&= h\left(\max\{B,0.5\}\right) . \label{eq:SWC_capacity_LB_temp3} 
\end{align}
The proof is complete by combining \eqref{eq:SWC_Capacity_UpperBound} and \eqref{eq:SWC_capacity_LB_temp3}.
\end{IEEEproof}

\subsection{Subblock Energy Constrained Codes}
Let $\mathcal{A}_{SEC}^{E_{max}}(B)$ denote the set of all feasible pairs $(L,w)$ which satisfy conditions \eqref{eq:SEC_OA_cond1} and \eqref{eq:SEC_OA_cond3}, such that corresponding $(L,w)$-SEC codes avoid outage. Then, the outage-constrained capacity for SWC codes, for energy buffer size $E_{max}$ and required energy per bit $B$, denoted $\mathcal{O}_{SEC}^{E_{max}}(B)$, is defined as
\begin{equation}
\mathcal{O}_{SEC}^{E_{max}}(B) \triangleq \max_{(L,w) \in \mathcal{A}_{SEC}^{E_{max}}(B)} C_{SEC}^{(L,w)} . \label{eq:Def_SEC_capacity_avoid_outage}
\end{equation}
If $E_{max} < 2B$ then condition \eqref{eq:SEC_OA_cond3} is satisfied only when $w=L$, and hence $\mathcal{O}_{SEC}^{E_{max}}(B) = 0$ in this case.  

The following proposition gives a lower bound on the outage-constrained capacity for SEC codes.
\begin{proposition} \label{prop:SEC_capacity_LB}
	For $E_{max} \ge 2B$, we have 
	\begin{equation}
	\mathcal{O}_{SEC}^{E_{max}}(B) \ge C_{SEC}^{(L,w)} , \label{eq:SEC_capacity_LB}
	\end{equation}
	where $\displaystyle L = \ceil{\frac{1}{(1-B)} \floor{\frac{E_{max}}{2B}} }$ and $w=\ceil{LB}$.
\end{proposition}
\begin{IEEEproof}
Conditions given by Thm.~\ref{thm:SEC_conditions} for avoiding outage are equivalent to the following properties:
\begin{enumerate}[(i)]
	\item Fraction of ones in $L$ length subblock is at least $B$.
	\item The number of zeros in each subblock is at most $\floor{E_{max}/(2B)}$.
\end{enumerate}
If $L = \ceil{\frac{1}{(1-B)} \floor{ \frac{E_{max}}{2B} } }$ and the number of ones in an $L$ length subblock is $\ceil{LB}$, then we can compute the number of zeros in the subblock as $L - \ceil{LB} = \floor{E_{max}/(2B)}$, thereby satisfying the above properties. Thus, the pair $(L,w)$ with $L = \ceil{\frac{1}{(1-B)} \floor{ \frac{E_{max}}{2B} } }$ and $w=\ceil{LB}$ belongs to the feasible set $\mathcal{A}_{SEC}^{E_{max}}(B)$, and the proof is complete using \eqref{eq:Def_SEC_capacity_avoid_outage}.
\end{IEEEproof}

The following upper bound on the outage-constrained capacity of SEC codes follows directly from the remark following Prop.~\ref{prop:SWC_Capacity_UpperBound}.
\begin{proposition} \label{prop:SEC_Capacity_UpperBound}
	We have 
	\begin{equation}
	\mathcal{O}_{SEC}^{E_{max}}(B) \le h\left(\max\{B,0.5\}\right) . \label{eq:SEC_Capacity_UpperBound}	
	\end{equation}
\end{proposition}

The next proposition shows that $\mathcal{O}_{SEC}^{E_{max}}(B)$ achieves the upper bound in~\eqref{eq:SEC_Capacity_UpperBound} when the receiver energy buffer is unbounded.
\begin{proposition}
	 We have
	\begin{equation}
	\lim_{E_{max} \to \infty} \mathcal{O}_{SEC}^{E_{max}}(B) = h\left(\max\{B,0.5\}\right) . \label{eq:SEC_capacity_large_L}
	\end{equation} 
\end{proposition}
\begin{IEEEproof}
Using \eqref{eq:SEC_Capacity_UpperBound}, we have $\lim_{E_{max} \to \infty} \mathcal{O}_{SEC}^{E_{max}}(B) \le h\left(\max\{B,0.5\}\right)$. Note from Prop.~\ref{prop:SEC_capacity_LB} that feasible subblock length $L \to \infty$ as $E_{max} \to \infty$. Hence, using \eqref{eq:SEC_capacity_LB} and \eqref{eq:SEC_Lw_capacity},
\begin{align*}
\lim_{E_{max} \to \infty} \mathcal{O}_{SEC}^{E_{max}}(B) &\ge \lim_{L \to \infty} \frac{1}{L} \log_2 \left[ \sum_{i=\ceil{LB}}^L \binom{L}{i} \right] \\
&= h\left(\max\{B,0.5\}\right) .
\end{align*}
\end{IEEEproof}

\subsection{Comparing outage-constrained capacities} \label{subsec:CompareOCC}
This subsection presents the main result of the paper that the outage-constrained capacity of RLL codes is just a lower bound to the outage-constrained capacity of SWC and SEC codes. 

\begin{theorem} \label{thm:SWC_RLL_capacity}
	We have
	\begin{equation}
\mathcal{O}_{SWC}^{E_{max}}(B) \ge \mathcal{O}_{RLL}^{E_{max}}(B). \nonumber 
	\end{equation}
	Moreover, the rate gap, $\mathcal{O}_{SWC}^{E_{max}}(B) - \mathcal{O}_{RLL}^{E_{max}}(B)$ is an increasing function of $E_{max}$.
\end{theorem}
\begin{IEEEproof}
When $E_{max} < B$, we have $\mathcal{O}_{SWC}^{E_{max}}(B) = \mathcal{O}_{RLL}^{E_{max}}(B)$. When $B \le E_{max} < 2B$, the conditions for avoiding outage using $(T,w)$-SWC codes (given by Thm.~\ref{thm:SWC_conditions}), imply that if a pair $(T,w)$ achieves capacity in \eqref{eq:SWC_Emax_B_capacity}, then we have $w=\ceil{TB}=T-1$. This, in turn, implies that $T-1 \ge \ceil{B/(1-B)}$. From \eqref{eq:SWC_Ineq}, we know that $C_{SWC}^{(T,T-1)}$ is a non-increasing function of $T$, and hence  $\mathcal{O}_{SWC}^{E_{max}}(B) = C_{SWC}^{(T,T-1)}$ with $T = 1 + \ceil{B/(1-B)}$. Combining this with Prop.~\ref{prop:RLL_SWC_eq} and Prop.~\ref{prop:RLL_capacity_compute} we get $\mathcal{O}_{SWC}^{E_{max}}(B) = \mathcal{O}_{RLL}^{E_{max}}(B)$ for $B \le E_{max} < 2B$. 

Now, an increase in $E_{max}$ only enlarges the set $\mathcal{A}_{SWC}^{E_{max}}(B)$ which leads to a potential increase in $\mathcal{O}_{SWC}^{E_{max}}(B)$ (see \eqref{eq:SWC_Emax_B_capacity}). On the other hand, $\mathcal{O}_{RLL}^{E_{max}}(B)$ does not vary with $E_{max}$ when $E_{max}$ exceeds $2B$, and hence $\mathcal{O}_{SWC}^{E_{max}}(B) - \mathcal{O}_{RLL}^{E_{max}}(B)$ is an increasing function of $E_{max}$. 
\end{IEEEproof}

\begin{theorem} \label{thm:SEC_RLL_capacity}
We have
\begin{equation}
\mathcal{O}_{SEC}^{E_{max}}(B) > \mathcal{O}_{RLL}^{E_{max}}(B) \mathrm{~~when~~} E_{max} \ge 2B . \nonumber
\end{equation}
Moreover, the rate gap $\mathcal{O}_{SEC}^{E_{max}}(B) - \mathcal{O}_{RLL}^{E_{max}}(B)$ is an increasing function of $E_{max}$.
\end{theorem}
\begin{IEEEproof}
We first show that if $E_{max} = 2B$, then $\mathcal{O}_{SEC}^{E_{max}}(B) > \mathcal{O}_{RLL}^{E_{max}}(B)$. Using \eqref{eq:SEC_OA_cond1}, \eqref{eq:SEC_OA_cond3} and $E_{max} = 2B$, we note that a SEC capacity achieving pair $(L,w)$ satisfies $w=\ceil{LB}=L-1$, which implies $L-1 \ge \ceil{B/(1-B)}$. Applying Lem.~\ref{lemma:SEC_capacity_one_zero}, we get
\begin{align}
\mathcal{O}_{SEC}^{E_{max}}(B) &= C_{SEC}^{(1+\ceil{B/(1-B)}, \ceil{B/(1-B)})} \nonumber \\
&\overset{(\mathrm{a})}{>} C_{SWC}^{(1+\ceil{B/(1-B)}, \ceil{B/(1-B)})} \nonumber \\
&\overset{(\mathrm{b})}{=} C_{RLL}^{(\ceil{B/(1-B)}, \infty)} \nonumber \\
&\overset{(\mathrm{c})}{=} \mathcal{O}_{RLL}^{E_{max}}(B) , \nonumber
\end{align}
where $(\mathrm{a})$ follows from Lem.~\ref{lemma:SWC_SEC_capacity}, $(\mathrm{b})$ follows from Prop.~\ref{prop:RLL_SWC_eq}, and $(\mathrm{c})$ follows from Prop.~\ref{prop:RLL_capacity_compute}. The proof is now complete by observing that $\mathcal{O}_{SEC}^{E_{max}}(B)$ is an increasing function of $E_{max}$ while $\mathcal{O}_{RLL}^{E_{max}}(B)$ does not vary with $E_{max}$ when $E_{max}$ exceeds $2B$.
\end{IEEEproof}

Theorem~\ref{thm:SEC_RLL_capacity} shows that the outage-constrained capacity of SEC codes is \emph{strictly} higher than that of RLL codes when $E_{max} \ge 2B$. The numerical results presented in the next section demonstrate that the outage-constrained capacity of SWC and SEC codes can be significantly higher than RLL codes. These results show that run-length limited codes are not the most suitable class of codes for simultaneous energy and information transfer.

\section{Numerical Results} \label{Sec:Numerical}

\begin{figure}[t]
	\centering
	\includegraphics[width=\myfigwidth\textwidth]{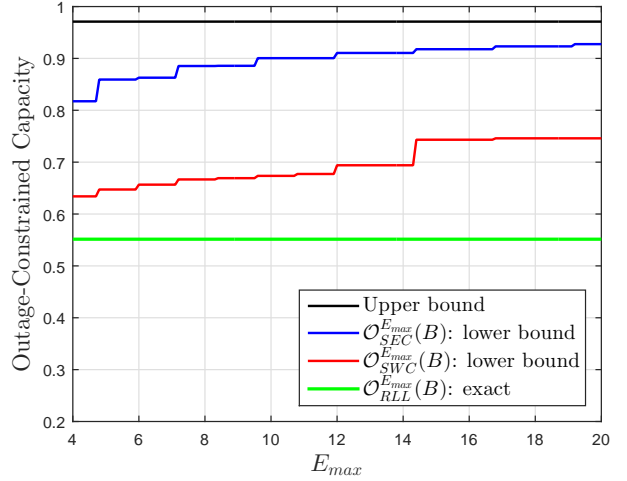}
	\caption{Constrained capacity comparison for $B=0.6$.}
	\label{Fig:OutageCapacity_versus_Emax}
\end{figure}

Fig.~\ref{Fig:OutageCapacity_versus_Emax} compares the outage-constrained capacity for different codes as a function of $E_{max}$ when $B=0.6$. The upper bound is computed using the expression $h\left(\max\{B,0.5\}\right)$ which corresponds to the code capacity when the fraction of ones in each codeword is at least $B$ (see Prop.~\ref{prop:SWC_Capacity_UpperBound} and the following remark). The lower bound for $\mathcal{O}_{SEC}^{E_{max}}(B)$  is computed using \eqref{eq:SEC_capacity_LB} and \eqref{eq:SEC_Lw_capacity}. As shown in Thm.~\ref{thm:SEC_RLL_capacity}, we observe from Fig.~\ref{Fig:OutageCapacity_versus_Emax} that $\mathcal{O}_{SEC}^{E_{max}}(B)$ is strictly greater than $\mathcal{O}_{RLL}^{E_{max}}(B)$, and the corresponding rate gap only increases with $E_{max}$. Further, we note from \eqref{eq:SEC_capacity_large_L} that $\mathcal{O}_{SEC}^{E_{max}}(B)$ tends to the line corresponding to the upper bound as $E_{max}$ tends to infinity. From Thm.~\ref{thm:SWC_RLL_capacity} we have $\mathcal{O}_{SWC}^{E_{max}}(B) \ge \mathcal{O}_{RLL}^{E_{max}}(B)$. The lower bound for $\mathcal{O}_{SWC}^{E_{max}}(B)$ is computed by combining Prop.~\ref{prop:SWC_capacity_LB} and Prop.~\ref{prop:SWC_capacity_LB_using_SEC_capacity2}. Using Prop.~\ref{prop:SEC_capacity_large_Emax} it follows that $\mathcal{O}_{SWC}^{E_{max}}(B)$ will eventually tend to the upper bound as $E_{max} \to \infty$. The exact value of $\mathcal{O}_{RLL}^{E_{max}}(B)$ is computed using Prop.~\ref{prop:RLL_capacity_compute}. Fig.~\ref{Fig:OutageCapacity_versus_Emax} demonstrates that the outage-constrained capacity of SWC and SEC codes is significantly higher than that of RLL codes. Further, the capacity gap only increases with increasing $E_{max}$.

\begin{figure}[!h]
	\centering
	\includegraphics[width=\myfigwidth\textwidth]{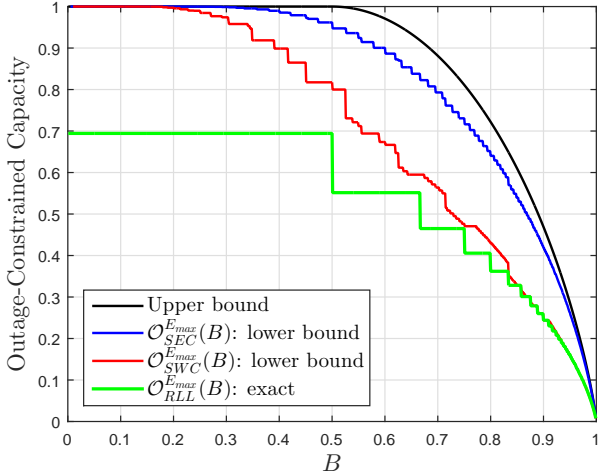}
	\caption{Constrained capacity comparison for $E_{max}=10$.}
	\label{Fig:OutageCapacity_versus_B}
\end{figure}
Fig.~\ref{Fig:OutageCapacity_versus_B} compares constrained capacity as a function of $B$ when $E_{max}=10$. The different curves in Fig.~\ref{Fig:OutageCapacity_versus_B} are obtained in a similar fashion as Fig.~\ref{Fig:OutageCapacity_versus_Emax}. Note that there are discrete jumps in Figs.~\ref{Fig:OutageCapacity_versus_Emax} and~\ref{Fig:OutageCapacity_versus_B} since the subblock length, the sliding window length, and corresponding weight constraints are all integer-valued. Fig.~\ref{Fig:OutageCapacity_versus_B} again shows that the outage-constrained capacity of SWC and SEC codes is much higher than that of RLL codes, especially for relatively small values of $B$. These figures demonstrate that run-length limited codes are not the most suitable class of constrained codes for providing simultaneous energy and information to an energy harvesting receiver.

\section{Reflections}

We compared the outage-constrained capacity of different code families under a common requirement of avoiding energy outage in simultaneous energy and information transfer systems. A key takeaway from the comparison is that while run-length limited (RLL) codes have been crucial in designing efficient codes for storage systems~\cite{ImminkBook91,Tang69,Franaszek70,Franaszek72,Siegel85,Siegel87,Zehavi88,Shamai90,ImminkBook99,Ghaffar91,Wijngaarden01,Yogesh06,Wijngaarden10,Cai17,Immink18}, these codes are \emph{not} well suited for simultaneous energy and information transfer. It was shown that the sliding window constrained (SWC) and the subblock energy constrained (SEC) codes provide significant rate improvement over RLL codes.

Over the years, different construction approaches with varied enhancements have been proposed for RLL codes~\cite{Ghaffar91,Wijngaarden01,Yogesh06,Wijngaarden10}. We remark here that SEC codes are also amenable to efficient implementation via concatenation~\cite{ForneyBook}, where the inner code is a \emph{heavy weight code}~\cite{Cohen10} and the outer code is a high rate code over large alphabet, such as a Reed-Solomon code~\cite{PetersonWeldon}.

The class of \emph{skip-sliding window (SSW) codes}, which generalize both SWC codes and SEC codes using a \emph{skip} value, was introduced in~\cite{Wu17_Arxiv,Wu18_ISIT}. As SWC and SEC codes are just special cases of the SSW code, it is expected that the outage-constrained capacity of the SSW code may strictly exceed that of SWC and SEC codes via an appropriate choice of the skip value. For given parameters of the SSW code, although the noiseless capacity can be computed by applying one of the many enumeration techniques presented in~\cite{Wu17_Arxiv}, the corresponding computational complexity may become prohibitively high for certain choice of parameters. An interesting area of future work is the efficient computation of the outage-constrained capacity of SSW codes. This involves combining the enumeration techniques for SSW codes in \cite{Wu17_Arxiv}, together with finding a feasible set of parameters for SSW code which avoid energy outage at the receiver.

\appendices
\section{Proof of Proposition \ref{prop:rll}}
\begin{IEEEproof}
	Note that we have $C_{RLL}^{(d+1,\infty)} \le C_{RLL}^{(d,\infty)}$ because any $(d+1,\infty)$-RLL sequence is also a $(d,\infty)$-RLL sequence. To prove $C_{RLL}^{(d+1,\infty)}$ is \emph{strictly} less than $C_{RLL}^{(d,\infty)}$, we note that $C_{RLL}^{(d,\infty)}$ is given by the base-two log of the largest real root of the polynomial $P_d(X) = X^{d+1} - X^d - 1$ (see \cite{Siegel87}), and any root of $P_d(X)$ cannot be a root of $P_{d+1}(X)$.
\end{IEEEproof}

\section{Proof of Proposition \ref{prop1}}
\begin{IEEEproof}
	For proving the lower bound on $C_{SWC}^{(T,w)}$, we will show that any $(T+m,w+m)$-SWC sequence is also a $(T,w)$-SWC sequence. Let $\mathbf{s}$ be a $(T+m,w+m)$-SWC sequence. Then a sliding window of length $T+m$ over $\mathbf{s}$ has at most $T-w$ zeros, which implies that a sliding window of length $T$ cannot have more than $T-w$ zeros, and hence $\mathbf{s}$ is also a $(T,w)$-SWC sequence. The upper bound in \eqref{eq:SWC_Ineq} can similarly be proved by showing that any $(T,w)$-SWC sequence is also a $(Tm,wm)$-SWC sequence.
\end{IEEEproof}

\section{Proof of Proposition \ref{prop:SWC_capacity_LB_using_SEC_capacity}}
\begin{IEEEproof}
	The upper bound on $C_{SWC}^{(T,w)}$ follows because any sequence satisfying the $(T,w)$-SWC constraint also satisfies the $(T,w)$-SEC constraint. Towards proving the lower bound, consider set $\mathcal{V}$ composed of binary vectors of length $T+w$, where the first $T$ bits of each vector have at least $w$ ones, while the last $w$ bits are all ones. Then any sequence formed by stacking vectors from $\mathcal{V}$ will satisfy the $(T,w)$-SWC constraint. The cardinality $|\mathcal{V}|$ is $\sum_{i=w}^T \binom{T}{i}$, and hence
	\begin{equation}
	M_{SWC}^{(T,w)}\left( (T+w)k \right) \ge \left[\sum_{i=w}^T \binom{T}{i}\right]^k . \nonumber
	\end{equation}
	Thus, $C_{SWC}^{(T,w)}$ satisfies
	\begin{align}
	C_{SWC}^{(T,w)} &= \lim_{k \to \infty} \frac{1}{k} \, \frac{\log M_{SWC}^{(T,w)}\left( (T+w)k \right)}{T+w} \nonumber \\
	&\ge \frac{\log \sum_{i=w}^T \binom{T}{i}}{T+w} = \frac{T}{T+w} C_{SEC}^{(T,w)} . \nonumber
	\end{align}
\end{IEEEproof}

\section{Proof of Proposition \ref{prop:SWC_capacity_LB_using_SEC_capacity2}}
\begin{IEEEproof}
	We first show that $C_{SWC}^{(T,w)} \ge C_{SEC}^{\left(T-1,\ceil{(T+w-2)/2}\right)}$. Let $\mathbf{s}$ be any $\left(T-1,\ceil{(T+w-2)/2}\right)$-SEC sequence. Consider a window of length $T$ sliding over $\mathbf{s}$. The number of zeros in this window are at most $z = 2(T-1 - \ceil{(T+w-2)/2})$, and so the number of ones in this window are at least $T-z = 2\ceil{(T+w-2)/2} -T + 2 \ge w$. This implies that $\mathbf{s}$ is also a $(T,w)$-SWC sequence, and hence $$C_{SWC}^{(T,w)} \ge C_{SEC}^{\left(T-1,\ceil{(T+w-2)/2}\right)}.$$
	
	We now show that $C_{SWC}^{(T,w)} \ge C_{SEC}^{\left(\floor{T/(m+1)},\ceil{w/m}\right)}$. Let $\mathbf{s}$ be any $\left(\floor{T/(m+1)},\ceil{w/m}\right)$-SEC sequence. Consider a window of length $(m+1)\floor{T/(m+1)}$ sliding over $\mathbf{s}$. This sliding window overlaps \emph{completely} with at least $m$ subblocks of length $\floor{T/(m+1)}$, and so the number of ones in the window are at least $m \ceil{w/m}$. This implies that $\mathbf{s}$ is a $\left( (m+1)\floor{T/(m+1)}, m \ceil{w/m} \right)$-SWC sequence, and
	\begin{align*}
	C_{SEC}^{\left(\floor{T/(m+1)},\ceil{w/m}\right)} &\le C_{SWC}^{\left( (m+1)\floor{T/(m+1)}, m \ceil{w/m} \right)} \\
	&\overset{(\mathrm{a})}{\le} C_{SWC}^{(T,w)} ,
	\end{align*}
	where $(\mathrm{a})$ follows from \eqref{eq:SWC_Ineq2}.
\end{IEEEproof}

\section{Proof of Theorem \ref{thm:swc}}
\begin{IEEEproof}
	We first note that if $\mathbf{s}$ is a $(d,\infty)$-RLL sequence, then it is also a $(d+1,d)$-SWC sequence. This claim follows because any zero in $\mathbf{s}$ is followed by at least $d$ ones, and hence a sliding window of length $d+1$ can only have a single zero.
	
	We next note that if $\tilde{\mathbf{s}}$ is a $(d+1,d)$-SWC sequence, then it is also a $(d,\infty)$-RLL sequence. This claim follows by observing that if a window of $d+1$ bits is placed over $\tilde{\mathbf{s}}$ in such a way that the starting bit in the window is a zero, then the following $d$ bits in the window following zero have to be all ones in order to satisfy the sliding window constraint.
\end{IEEEproof}

\section{Proof of Lemma \ref{lemma:SWC_SEC_capacity}}
\begin{IEEEproof}
	From Prop.~\ref{prop:RLL_SWC_eq}, $C_{SWC}^{(T,T-1)}$ is equal to $C_{RLL}^{(T-1,\infty)}$, which is given by the logarithm of the largest real root of $f(X) = X^T - X^{T-1} - 1$ \cite{Siegel87}. Using Descartes' rule of signs~\cite{Wang04_DROS}, and the facts $f(1) <0$ and $f(2) > 0$ for $T > 1$, it follows that $f$ only has one positive root. Let $\alpha > 1$ denote the positive root of $f$ and let $\beta > 1$ be another given positive number. Then $\beta > \alpha$ if and only if $f(\beta) > 0$. Thus, if we let $\beta = (T+1)^{1/T}$, then $ C_{SEC}^{(T,T-1)} = \log_2 \beta$ and the proof is complete if we show that $f(\beta)>0$. Now, we have
	\begin{align}
	f(\beta) > 0 &\iff T > (T+1)^{(T-1)/T} \nonumber \\
	&\iff T+1 > \left(1 + \frac{1}{T}\right)^T \nonumber \\
	&\iff T+1 > \sum_{i=0}^T \binom{T}{i} \frac{1}{T^i} \nonumber 
	\end{align}
	where the last inequality is true as $\binom{T}{i} \frac{1}{T^i} < 1$ for $1 < i \le T$, thereby proving $C_{SEC}^{(T,T-1)} = \log_2 \beta > \log_2 \alpha = C_{SWC}^{(T,T-1)}$.
\end{IEEEproof}

\section{Proof of Lemma \ref{lemma:SEC_capacity_one_zero}}
\begin{IEEEproof}
	We have
	\begin{align}
	&~~~~~~~~~ C_{SEC}^{(T,T-1)} > C_{SEC}^{(T+1,T)} \nonumber \\
	&\iff (T+1)^{1/T} > (T+2)^{1/(T+1)} \nonumber \\
	&\iff T+1 > \left(1 + \frac{1}{T+1}\right)^T , \nonumber
	\end{align}
	where the last inequality follows from the fact that $\binom{T}{i} \frac{1}{(T+1)^i} < 1$ for $1 \le i \le T$.
\end{IEEEproof}

\balance



\end{document}